\documentclass[prb,twocolumn]{revtex4}
\usepackage{graphicx}
\usepackage{dcolumn}
\usepackage{bm}
\usepackage[english]{babel}

\begin{document}

\preprint{}

\title{Excitonic relaxation and transfer in semiconductor nanocrystals}

\author{V.A. Burdov}

\affiliation{Lobachevsky State University of Nizhny Novgorod, 23 Gagarin avenue, 603950 Nizhny Novgorod, Russian Federation}

\date{\today}

\begin{abstract}

Some basic radiative and non-radiative processes taking place in semiconductor nanocrystals are discussed, and rates of these processes are calculated. In particular, in the present review we explore both intra-crystallite processes, such as the photon emission, Auger recombination, phonon-assisted exciton relaxation, capture on surface defects, multi-exciton generation, and inter-crystallite processes realized through the exciton migration in ensembles of nanocrystals. Exciton transitions, creation, and annihilation accompanied by the photon and (or) phonon emission or absorption are considered from the point of view of their influence on the optical properties of nanocrystals.

\end{abstract}

\maketitle

\section{Introduction}

As known, semiconductor nanostructures form a basis for modern electronic technologies. In particular, they are employed in a wide range of optical applications, such as optoelectronics, photonics, photovoltaics, bio-sensing, \textit{etc}. It is a very important property of nanostructured materials that their electronic spectra are completely, or partially, discrete and depend on the nanostructure size. This property is most pronounced for zero-dimensional (0D) objects---nanocrystals---in which electron motion is fully localized in all directions. Consequently, at least a part of the nanocrystal energy spectrum is completely discrete. In the limiting case of strong quantum confinement effect, when the nanocrystal size is much less than the effective exciton Bohr radius, the electron and hole energies strongly depend on the nanocrystal size, which leads to the size-dependent photon frequency equal to the energy of the electron-hole radiative transition. This allows one to control the frequency (or wavelength) of light emitted (absorbed) by the nanocrystals.

Different methods can be used for preparation of crystallites with sizes not greater than several nanometers: ion implantation;~\cite{Kanemitsu,Zhuravlev,NIMB} chemical vapor deposition from a gas phase;~\cite{Negro} magnetron sputtering;~\cite{Tsybeskov} colloidal synthesis;~\cite{Nayfeh} electron beam epitaxy;~\cite{Meldrum} \textit{etc}. One of these methods---growing in solutions---allows to obtain nanocrystals of almost spherical shape with very narrow size distribution and, as a consequence, with very weak dispersion of the photon frequencies, which is, as a rule, extremely desirable for applications. This method was first suggested for A$_2$B$_6$ nanocrystals by Murray \textit{et al.}~\cite{Murray} and widely employed later (see, e.g., papers of Kudera \textit{et al.}, Reiss, Gaponik and Rogach~\cite{book1} for review).

In our present study, we mainly concentrate on radiative electron-hole transitions which provide light emission in visible range. However, light emission is a result of strong concurrence with various non-radiative processes affecting the photoluminescence, such as Auger recombination, capture on dangling bonds, multi-phonon intra-band relaxation, and some others. Detailed discussion of all these processes is given, e.g., in the book of Delerue and Lannoo.~\cite{book} Besides, in real systems, as a rule, one has to deal not with isolated nanocrystals but rather with their ensembles in which nonradiative energy exchange between the nanocrystals is possible \textit{via} tunnel or F\"{o}rster~\cite{Forster} exciton migration. Previous experimental investigations of multilayer SiO$_x$/SiO$_2$ structures,~\cite{Linnros,Heitmann,Glover} porous Si,~\cite{Ben} three-dimensional (3D) ensembles of Si~\cite{Priolo,Balberg} and A$_2$B$_6$~\cite{Kagan,Crooker,Yu,Kawazoe} crystallites demonstrated a dependence of their optical properties on the nanocrystal density and spatial arrangement, which was treated as a migration's manifestation.

In the past decade, growing interest is attracted also to the multi-exciton dynamics in nanocrystals in view of their high potential in photovoltaic applications. Such processes as carrier multiplication (or, multi-exciton generation) in nanocrystals, as well as the Auger recombination, that is a fast reverse process with respect to the carrier multiplication influencing the latter, are widely discussed.~\cite{Klimov,Bruhn,Haverkort} In the present review, we also touch upon some aspects of the multi-exciton generation in crystallites.

Special attention is traditionally paid to nanostructured silicon.~\cite{Pavesi,Khriach,Ray,Barba,Priolo1} Its widest use in microelectronics, high purity, natural abundance, low cost, and non-toxicity make this material highly attractive also for various applications (other than the microelectronics) including some optical ones. Below, we shall discuss some cases of the exciton-photon interactions, as well as main non-radiative processes, mentioned above, in Si nanocrystals and other nanostructured materials.

The basic quantitative characteristic of any process is its rate---probability per unit time or reciprocal lifetime. Precisely this rate will determine a relative efficiency of any process among all possible processes taking place in the system. Therefore, in the following we shall focus on the rates and efficiencies of the radiative and competing with them non-radiative processes, and begin with the radiative inter-band recombination. Further, we compute the rates of all the considered processes with the Fermi golden rule that is usually used for these purposes.

\section{Radiative transitions}

The radiative decay rate for some no-phonon inter-band electron-hole transition (exciton annihilation) can be found as (after summation over all photon states involved in the transition)
\begin{equation}
\tau_{R}^{-1} = \frac{4e^2\varepsilon_{if}|{\bf p}_{if}|^2\kappa\sqrt{\epsilon_{out}}}{3m_0^2\hbar^2c^3}.
\end{equation}
Here, $m_0$ and $-e$ are the free electron mass and charge, respectively, $c$ is the speed of light, and $\hbar$ denotes the Planck constant. $\varepsilon_{if}$ and ${\bf p}_{if}$ denote the transition energy between the initial and final single-electron states and the momentum matrix element, respectively. The parameter $\kappa$ arises due to the difference in the permittivities of the crystallite ($\epsilon$) and its surrounding ($\epsilon_{out}$). Hereafter, we consider the crystallite as a sphere with radius $R$ and permittivity $\epsilon$ embedded in a homogeneous medium with bulk permittivity $\epsilon_{out}$. In this case, $\kappa$ is defined as~\cite{Thranhardt}
\begin{equation}
\kappa = \frac{9\epsilon_{out}^2}{(2\epsilon_{out}+\epsilon)^2}.
\end{equation}
$\kappa$ varies within a wide range depending on the dielectric constants of both crystallite and host matrix.

\subsection{Nanocrystals formed of direct-band-gap materials}

Majority of semiconductor A$_3$B$_5$, A$_2$B$_6$, A$_4$B$_6$ materials has, as a rule, direct gap. In this case evaluation of the momentum matrix element ${\bf p}_{if}$, that has to be done for calculation of $\tau_{R}^{-1}$, is not a difficult problem. Within the framework of the envelope function approximation ${\bf p}_{if}$ can be written as
\begin{equation}
{\bf p}_{if} = {\bf p}_{cv}\langle F_{c,i}|F_{v,f}\rangle,
\end{equation}
where ${\bf p}_{cv}$ is a standard bulk-like optical matrix element of the inter-band transition, while $F_{c,i}$ and $F_{v,f}$ stand for the envelope functions of the initial state in the conduction band and the final state in the valence band (in case of the photon emission), respectively. The scalar product of the envelope functions determines the selection rules for the transition.

In a simplest case, when the nanocrystal is treated as an infinitely deep spherical potential well with radius $R$, the electron and hole ground states, which are supposed to be initial and final states for the radiative transition, are described by the same envelope functions. Consequently, their scalar product equals unity, and the radiative recombination rate turns out to be almost independent of the crystallite radius $R$. This dependence manifests only in $\varepsilon_{if}=\varepsilon_g(R)$ for the electron-hole transition. In the limiting case of a strong quantum confinement regime, when $R\ll a_B^*$, where $a_B^*$ stands for the effective Bohr radius of a bulk-like exciton, the model of infinitely deep spherical potential well yields: $\varepsilon_g(R) = \varepsilon_g(\infty) + \hbar^2\pi^2/2mR^2$, where $\varepsilon_g(\infty)$ is the gap of the bulk semiconductor, and $m$ is the reduced effective mass of the electron-hole pair.

Note that nearly constant dependence of $\tau_{R}^{-1}$ on the crystallite size takes place only for the crystallites with the direct-gap electronic structure. As we shall see below, for silicon nanocrystals this statement is absolutely incorrect---Si nanocrystals demonstrate very strong size-dependence of the radiative recombination rates, especially, for the transitions with no phonon assistance.

It is now possible to estimate $\tau_{R}^{-1}$ for a crystallite formed of some typical direct-band-gap semiconductor, e.g. GaAs, and embedded in a host matrix like, e.g. AlAs. Taking for estimations $|{\bf p}_{cv}|\sim 2\pi\hbar/a$, with $a$ being the lattice constant, and substituting static bulk values of dielectric constants of GaAs ($\epsilon =12.5$) and AlAs ($\epsilon_{out}=11$) into $\kappa$, one obtains $\kappa\sqrt{\epsilon_{out}}\sim 3$. As a result, the radiative recombination rate is as follows:
\begin{equation}
\tau_{R}^{-1}\sim 10^9\varepsilon_g(R).
\end{equation}
Here, the rate value is calculated in s$^{-1}$ if the energy gap is taken in eV. Since $\varepsilon_g(R)\sim$ a few electron-Volts, the radiative rate is about 10$^9$ s$^{-1}$. This theoretical estimation has multiple experimental confirmations.

Such estimations of the rate values remain valid in a wide range of temperatures---from room temperature to $\sim$ helium one. However at low temperatures the rate sharply decreases. This decrease is due to a relatively weak electron-hole exchange and spin-orbit interactions which split the eight-fold (in a spherical crystallite) exciton ground state into the states with different total angular momentum: $I = 1$ and $I = 2$, where the total excitonic angular momentum ${\bf I} = {\bf J} + {\bf s}$ with ${\bf J}$ and ${\bf s}$ being the hole angular momentum ($J = 3/2$) and the electron spin ($s = 1/2$), respectively. As described by Delerue and Lannoo,~\cite{book} in the strong confinement regime the five-fold degenerate level ($I = 2$) has a lowest energy corresponding to a ground excitonic state. However this exciton is ``dark''; it may not emit a photon. The exciton with higher, triply degenerate, energy and $I = 1$ is ``bright'', but at low temperature, this level is not, in fact, occupied. As a result, the radiative transition does not occur, which means a strong decrease of the recombination rate in this case. A similar effect takes place also for nanocrystals of indirect-gap semiconductors.

\subsection{Silicon nanocrystals}

During the past decades, Si crystallites were extensively studied from the point of view of their use in optical applications.~\cite{Pavesi,Khriach,Ray,Barba,Priolo1} In order to improve their optical properties, various physical and chemical methods have been used: the introduction of shallow impurities;~\cite{Tetel,Fujii,JPCM,Nomoto,Klimesova} formation in different matrices,~\cite{Kim,Klangsin} plasma,~\cite{Nozaki,Zhou} and colloidal solutions.~\cite{Baldwin,Wolf,Bell,Zhou1,Carroll} As a result, the theoretical predictions or experimental observations of enhanced photoluminescence intensity,~\cite{Tetel,Fujii,JPCM,Nomoto} radiative recombination rates,~\cite{Klimesova,PRB,Kusova,Ma,Ma1,Wang,Light,Poddubny,JPCC,JAP1} or quantum efficiency of photon generation~\cite{Sangg} have been reported.

Traditionally, doping is considered as an effective means for modification of the electronic properties of bulk Si as well as of Si crystallites.~\cite{PLA,NRL,JPCM08,Chelik,Oliva,Arduca,Nomoto1} Such a modification, in turn, influences the electron-hole radiative transitions. It was revealed that doping of Si nanocrystals with P or Li is (under certain conditions) capable of improving their emittance.~\cite{Fujii,JPCM,Klimesova,Yang} In particular, the performed calculations~\cite{JPCM,PRB,JPCC,JAP1} show that doping with P or Li can essentially increase the radiative rates.

For nanocrystals with $R\gtrsim 1$ nm some semi-empirical methods are usually employed for computing their electronic structure, such as envelope function approximation~\cite{Hybertsen,Yassi,AOT} or tight binding model~\cite{Delerue}. The performed calculations of the phonon-assisted radiative recombination rates in Si crystallites in SiO$_2$ matrix yielded the values varying from $\sim 10^2$ to $\sim 10^5$ s$^{−1}$ as the crystallite radius decreases from 3 to 1 nm. Within a simplest model of an infinitely deep spherical potential well, the dependence of the rate on the crystallite radius is  $\tau_{R}^{-1}\sim R^{-3}$.~\cite{AOT} For the no-phonon radiative transitions the rates turn out to be much less: from one to three orders of magnitude with increasing sizes within the same size range. In this case, rate $\tau_{R}^{-1}\sim R^{-8}$,~\cite{AOT} i.e. sharply drops as $R$ increases. At the same time, doping with phosphorus provides several orders of magnitude increase of the radiative recombination rates for the no-phonon transitions. The rates become even 1 - 3 orders (with increasing sizes) greater than the rates of the phonon-assisted transitions and slowly decrease with increasing radius, especially, at high concentration of phosphorus.~\cite{PRB}

\begin{figure}[b]
  \centering
  \includegraphics[scale=0.9]{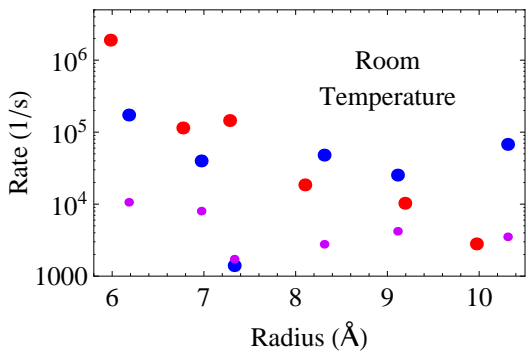}
  \caption{Rates of radiative transitions for undoped (small violet dots), P-doped (big blue dots), and Li-doped (big red dots) Si nanocrystals listed in Table 1.} \label{figure1}
\end{figure}

As the calculations show, the radiative transitions (both no-phonon and phonon-assisted) become faster in small Si crystallites. However for crystallites with $R\lesssim 1$ nm all the semi-empirical methods of calculation turn out to be essentially less accurate. In this case, first-principle calculations have to be carried out. In Fig. 1 the radiative recombination rates computed at room temperature $T$ within the Casida's version~\cite{Casida} of the time-dependent density functional theory (TDDFT) are depicted for P- and Li-doped, and undoped (for comparison) H-passivated Si nanocrystals, whose chemical formulae and radii are presented in Table 1. Here, vacuum was chosen as an environment; therefore $\epsilon_{out}=1$, while $\epsilon$ was chosen equal to 12. In this case, the rate values become $\sim 16$ times less compared to the case of Si nanocrystals embedded in silicon dioxide due to the factor $\kappa\sqrt{\epsilon_{out}}$.

\vspace{0.25 cm}
\noindent {\bf Table 1.} Radii $R$ ({\AA}) of the undoped and doped with P or Li hydrogen-coated silicon nanocrystals.

\begin{tabular}[c]{cccccc}
  \hline
   & $R$ &  & $R$ &  & $R$ \\ \hline
  Si$_{47}$H$_{60}$ & 6.18 & Si$_{46}$H$_{60}$P & 6.17 & Si$_{42}$H$_{64}$Li & 5.98 \\
  Si$_{71}$H$_{84}$ & 6.97 & Si$_{70}$H$_{84}$P & 6.97 & Si$_{66}$H$_{64}$Li & 6.77 \\
  Si$_{87}$H$_{76}$ & 7.33 & Si$_{86}$H$_{76}$P & 7.33 & Si$_{82}$H$_{72}$Li & 7.28 \\
  Si$_{123}$H$_{100}$ & 8.31 & Si$_{122}$H$_{100}$P & 8.31 & Si$_{106}$H$_{120}$Li & 8.10 \\
  Si$_{167}$H$_{124}$ & 9.11 & Si$_{166}$H$_{124}$P & 9.10 & Si$_{172}$H$_{120}$Li & 9.19 \\
  Si$_{239}$H$_{196}$ & 10.31 & Si$_{238}$H$_{196}$P & 10.30 & Si$_{220}$H$_{144}$Li & 9.97 \\
  \hline
\end{tabular}

\vspace{0.25 cm}

As seen, on the whole, the doped crystallites demonstrate one-two orders of magnitude faster transitions than the undoped ones. For doped with phosphorus Si crystallites, increasing rates of the radiative transitions have been explained by the efficient $\Gamma$-X mixing of the electronic states caused by the short-range field of the phosphorus ion.~\cite{JPCM,PRB} Meanwhile, the radiative decay in the Li-doped crystallites substantially quickens due to the high density of the conduction states in the nearest vicinity of the inter-band energy gap,~\cite{JAP1} for which $\hbar\omega - \varepsilon_g\lesssim k_BT$, where $k_B$ is the Boltzmann constant.

Surface chemistry is another method capable of modifying electronic structure of nanocrystals. For small nanocrystals, with sizes of the order of a nanometer, this method is especially efficient because of rising surface-to-volume ratio at decreasing size. It has been shown earlier both experimentally and theoretically that chemical synthesis allows to fabricate Si nanocrystals with various ligand coatings, which emit light in visible range with $\sim$ ns typical radiative lifetimes. For instance, Si nanocrystals with alkyl (CH$_3$-based) passivation demonstrate~\cite{Light} fast no-phonon radiative transitions with lifetimes about 10 ns in contrast to nanocrystals with (partially) oxydized surface, where the lifetimes range from milli- to microseconds. Theoretical model, based on the tight-binding calculations,~\cite{Poddubny} explains this effect by the formation of direct-like electronic structure due to the CH$_3$-capping of the crystallites.

Let us now consider some other kind of the surface reconstruction of the crystallites---halogenation. In Table 2 we show the calculated rates of radiative transitions between the highest occupied molecular orbital (HOMO) and the lowest unoccupied molecular orbital (LUMO) in completely halogenated with Cl, Br, and H silicon crystallites.

As seen from the table, a chlorine coating, in the mean, leads to a certain rate reduction with respect to the case of a hydrogen coating. Passivation with bromine strengthens the reduction, which is a consequence of a stronger separation of the electron and hole densities in the Br-passivated crystallites.~\cite{JLett,PCCP}

\vspace{0.25 cm}
\noindent {\bf Table 2.} Rates (s$^{-1}$) of the radiative HOMO-LUMO transitions in halogenated silicon nanocrystals.

\begin{tabular}[c]{cccc}
  \hline
   & X = H & X = Cl & X = Br \\ \hline
  Si$_{35}$X$_{36}$ & $3.7\times 10^5$ & 3.3 & 2.7 \\
  Si$_{59}$X$_{60}$ & $2.9\times 10^4$ & $4.2\times 10^4$ & $3.0\times 10^2$ \\
  Si$_{87}$X$_{76}$ & $1.6\times 10^3$ & $2.4\times 10^2$ & 36 \\
  Si$_{123}$X$_{100}$ & $3.6\times 10^3$ & $4.2\times 10^3$ & $1.7\times 10^5$ \\
  Si$_{147}$X$_{100}$ & $2.2\times 10^4$ & $4.3\times 10^2$ & 15 \\
  Si$_{163}$X$_{116}$ & $6.1\times 10^3$ & $3.0\times 10^4$ & $3.1\times 10^3$ \\
  Si$_{175}$X$_{116}$ & $3.5\times 10^4$ & $1.3\times 10^2$ & $3.5\times 10^2$ \\
  Si$_{317}$X$_{172}$ & $1.4\times 10^4$ & $2.4\times 10^3$ & 63 \\ \hline
\end{tabular}

\vspace{0.25 cm}

As has been shown earlier~\cite{JLett,PCCP,JETP} this trend takes place not only for the HOMO-LUMO transitions, but also for the transitions with higher energies. It is important to emphasize that even for greater photon energies, the rates in H-passivated crystallites do not exceed $10^8$ s$^{-1}$, while for Cl- and Br-passivated crystallites, this limiting value is one-four orders of magnitude smaller. Therefore, it is possible to conclude that the complete halogen passivation (see also the work of Ma et al.~\cite{Ma1} for F-passivated Si nanocrystals) significantly slows down the inter-band radiative recombination in Si nanocrystals.

\section{Nonradiative transitions}

It is known that an emittance of a nanocrystals' ensemble depends not only on the intensity of the radiative transitions but rather on the interplay of both radiative and nonradiative processes. Among latter, one can single out Auger recombination, capture on dangling bonds, and cooling of hot carriers. Below, we briefly discuss all these processes.

\subsection{Auger recombination}

There are two simplest electron-hole configurations sufficient for realization of Auger recombination: two electrons + one hole (negative trion); or one electron + two holes (positive trion). These trion states may transform into a highly excited electron or hole, respectively, \textit{via} Auger \textit{eeh} or \textit{ehh} processes. Below, we consider both processes formally assuming a negative or positive trion to be created in a crystallite, as shown schematically in Fig. 2.

The rate of the Auger transition is as follows:
\begin{equation}
\tau_{A}^{-1} = \frac{2\Gamma}{\hbar}\sum_f\frac{|U_{if}|^2}{\Gamma^2+(\varepsilon_f-\varepsilon_g)^2},
\end{equation}
where $U_{if}$ is the matrix element of the two-particle Coulomb interaction operator, the two-electron wave functions are the products of the single-electron wave functions, and the delta function was broadened with the half-width $\Gamma=10$ meV. All the possible final states of the Auger electron with energy $\varepsilon_f$ (shown in Fig. 2) are summed up. The initial states were chosen so that electrons and holes are in the LUMO and HOMO states, respectively.

\begin{figure}
  \centering
  \includegraphics[scale=0.9]{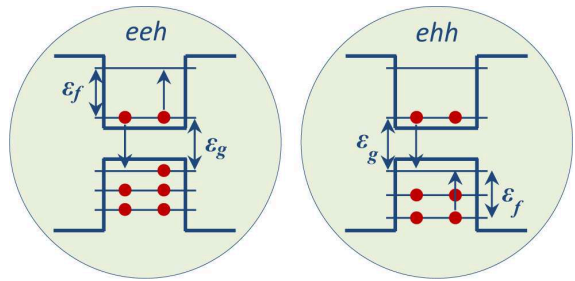}
  \caption{\textit{eeh} and \textit{ehh} Auger recombination in nanocrystals. Arrows indicate Auger transitions from the initial to the final states.} \label{figure2}
\end{figure}

Usually, Auger recombination is an efficient process that can ``shunt'' other photon-assisted processes such as light emission or multiple exciton generation. For instance, in Si crystallites, the Auger rates can be sufficiently high, as the measurements~\cite{Stolle,Pevere} and calculations~\cite{JPCC,Sevik,Delerue1,Mahdo,Semi} show. Therefore, weakening of the Auger process is an actual problem. As turned out, coating with halogens can slow down the Auger transitions.

Below, we compute the Auger rates and demonstrate their reduction for Si crystallites due to the surface halogenation. We take into account both the long-range ($U_0({\bf r}_1,{\bf r}_2$)) and the short-range ($U_1({\bf r}_1,{\bf r}_2$)) parts of the Coulomb interaction. The first one is a macroscopic electrostatic field modified by the nanocrystal boundary
\begin{eqnarray}
U_0({\bf r}_1,{\bf r}_2) & = & \frac{e^2}{\epsilon |{\bf r}_1-{\bf r}_2|}
\nonumber \\
& & + \frac{e^2(\epsilon - 1)}{\epsilon R} \sum_{l=0}^{\infty}\frac{(l+1)P_{l}(\cos\theta)}{l\epsilon+l+1}\frac{r_1^{l}r_2^{l}}{R^{2l}},
\end{eqnarray}
while the second one describes at a microscopic level the point charges interaction at short distances:
\begin{equation}
U_1(r) = \frac{e^2}{r}[Ae^{-\alpha r}+(1-A-1/\epsilon)e^{-\beta r}].
\end{equation}
Here $r=|{\bf r}_1-{\bf r}_2|$, $A=1.142$, $\alpha = 0.82/a_B$, and $\beta = 5/a_B$,~\cite{PRB} where $a_B$ stands for the Bohr radius, and we set $\epsilon_{out}=1$ as before. The calculated values of $\tau_{A}^{-1}$ are listed in Table 3 for both \textit{eeh} and \textit{ehh} trion annihilation in the Si$_n$X$_m$ crystallites considered here.

We can see that the halogen coating, as a rule, considerably slows down the Auger recombination as compared to that with a hydrogen coating. For quantitative estimation of the rate reduction, we have averaged the decimal logarithms of $\tau_{A}^{-1}$ for all three types of crystallite coatings, and obtained for the \textit{eeh} process: $\langle\lg(\tau_{A(\text{H})}^{-1})\rangle =11.66$; $\langle\lg(\tau_{A(\text{Cl})}^{-1})\rangle =10.55$; and $\langle\lg(\tau_{A(\text{Br})}^{-1})\rangle =10.60$. Similar values for the \textit{ehh} process are as follows: $\langle\lg(\tau_{A(\text{H})}^{-1})\rangle =11.86$; $\langle\lg(\tau_{A(\text{Cl})}^{-1})\rangle =11.44$; and $\langle\lg(\tau_{A(\text{Br})}^{-1})\rangle =11.20$. Hence, on an average, the \textit{eeh} Auger rate becomes an order of magnitude less because of the halogenation, while reduction of the \textit{ehh} rate turns out to be smaller. Similar suppression of Auger recombination caused by the surface modification was also recently predicted for GaSb crystallites.~\cite{Calif}

\vspace{0.25 cm}
\noindent {\bf Table 3.} Auger rates ($\times 10^{10}$ s$^{-1}$) for \textit{eeh} and \textit{ehh} processes in halogenated silicon nanocrystals.

\begin{tabular}[c]{ccccccc}
  \hline
   & X=H & X=Cl & X=Br & X=H & X=Cl & X=Br \\
   & ({\it eeh}) & ({\it eeh}) & ({\it eeh}) & ({\it ehh}) & ({\it ehh}) & ({\it ehh}) \\ \hline
  Si$_{35}$X$_{36}$ & 390 & 3 & 4 & 40 & 163 & 10 \\
  Si$_{59}$X$_{60}$ & 78 & 0.3 & 7 & 5 & 27 & 19 \\
  Si$_{87}$X$_{76}$ & 120 & 4 & 2 & 1780 & 59 & 11 \\
  Si$_{123}$X$_{100}$ & 8 & 3 & 280 & 108 & 25 & 24 \\
  Si$_{147}$X$_{100}$ & 59 & 84 & 2 & 474 & 20 & 65 \\
  Si$_{163}$X$_{116}$ & 120 & 21 & 100 & 47 & 9 & 19 \\
  Si$_{175}$X$_{116}$ & 270 & 2 & 2 & 161 & 90 & 0.8 \\
  Si$_{317}$X$_{172}$ & 0.3 & 0.4 & 0.02 & 5 & 3 & 78 \\ \hline
\end{tabular}

\vspace{0.25 cm}

It is also seen from Table 3, that the rate values are widely distributed. This is mainly due to the resonant profile of the rate [Eq. (5)] predicted earlier theoretically for Si,~\cite{Semi} A$_3$B$_5$,~\cite{Efros} and A$_2$B$_6$~\cite{Efros1}  crystallites. The resonant profile is caused by the discreteness of the energy spectrum (and, therefore, $\varepsilon_f$) in the crystallites. If the energy $\varepsilon_f$ of some excited state coincides (or almost coincides) with $\varepsilon_g$, then the rate becomes maximal owing to one resonant term in the sum in Eq. (5). If $\varepsilon_g$ falls between the levels, the rate decreases because of the absence of resonant terms in the sum. Varying the crystallite size, one can change both $\varepsilon_g$ and $\varepsilon_f$. Therefore, it is possible to transfer the system from a resonant to a nonresonant regime and \textit{vice versa}.

\subsection{Capture on dangling bonds}

There is one more relatively fast process inhibiting efficient light emission from nanocrystals---capture on surface defects called P$_{b}$-centers or dangling bonds. The P$_b$-centers produce reasonably deep energy levels within the nanocrystal gap. The capture on neutral dangling bonds is a multi-phonon process, and its rate strongly depends on temperature. The earlier performed calculations~\cite{book} for Si crystallites yielded $\tau_{C}^{-1}$ shown in Fig. 3 as function of the nanocrystal radius in comparison with the rates of the Auger recombination and radiative transitions.

It is seen that at the nanocrystal radius greater than $\sim 1$ nm, the capture rate becomes greater than the radiative recombination rate, and beginning with the radii close to 1.5 nm the capture on dangling bond turns out to be even faster than the Auger recombination. Obviously, for nanocrystals whose radii are greater than 1.5 nm the nonradiative trapping on the surface defects suppresses all the other competitive relaxation processes inside the nanocrystals if the surface dangling bonds are unpassivated.

\begin{figure}
  \centering
  \includegraphics[scale=0.9]{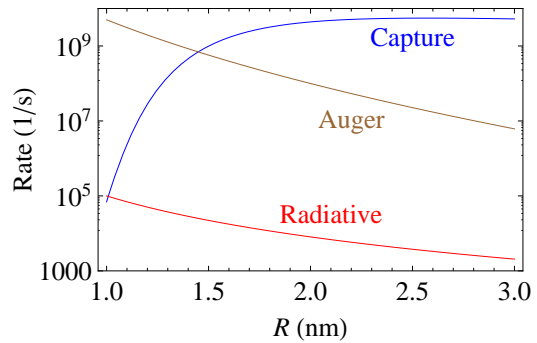}
  \caption{Rates of the capture on dangling bonds compared to the rates of radiative phonon-assisted transitions and Auger recombination in Si nanocrystals.} \label{figure3}
\end{figure}

\subsection{Phonon-assisted relaxation of hot carriers}

It is well known that highly excited (``hot'') electrons, holes, or excitons may relax to the lower states in order to minimize the system energy. Such a relaxation is accompanied by the phonon emission due to the interaction between electrons and lattice vibrations. For example, within the framework of the rigid-ion model, the electron-phonon interaction is described by the following operator:
\begin{equation}
{\hat W} = -\sum_{{\bf n},s}{\hat {\bf u}}_{{\bf n}s}\nabla V_{at}({\bf r}-{\bf R}_{{\bf n}s}).
\end{equation}
Here, $V_{at}({\bf r}-{\bf R}_{{\bf n}s})$ stands for the electron potential energy in the field of the $s$-th atom situated in the ${\bf n}$-th primitive cell of the lattice. ${\bf R}_{{\bf n}s}$ is the position-vector of this atom in equilibrium, while ${\hat {\bf u}}_{{\bf n}s}$ is the atomic displacement operator that depends on the phonon creation and annihilation operators. The rigid-ion model is more suitable when describing the electron-phonon interaction in multi-valley semiconductors, such as Si and Ge, where the processes of inter-valley scattering become often important.

In nanocrystals, because of a discreteness of their energy spectrum, the rate of the single-phonon emission strongly depends on the electron transition energy~\cite{Inoshita,book} that is defined by the level spacing. In particular, if the spacing between two adjacent levels turns out to be greater than the phonon energy, then such an electron transition becomes forbidden, and no phonons are emitted in this case. In the opposite case, the phonon-assisted electron transition is allowed. As a rule, the energy spectrum in nanocrystals becomes sparse upon approaching the energy gap. Therefore, the single-phonon relaxation stops at some energy level, and the relaxing carriers accumulate above (electrons) or below (holes) this level. This effect is often referred to as a phonon bottleneck.~\cite{Bastard} It does not mean, however, that the energy minimum may not be achieved. Simply, beginning with this moment, the relaxation transforms into a multi-phonon one. Usually, multi-phonon processes are slower compared to the single-phonon transitions. Therefore, the bottleneck effect does not disappear completely.

Estimations of the single-phonon relaxation rates in Si nanocrystals~\cite{Prokofiev,YassiAPL} yield typical values of the order of $10^{10}-10^{12}$ s$^{-1}$. At the same time, the multi-phonon relaxation rates calculated within the Huang-Rhys model~\cite{book,YassiPE,Moskal} for various transitions vary from $10^7$ s$^{-1}$ to $10^{11}$ s$^{-1}$ for nanocrystals whose diameters do not exceed 5 nm. It is worth to note that the rates sharply reduce as the nanocrystal diameter decreases. Similar values of the multi-phonon relaxation rates was obtained experimentally for 15 nm self-assembled InGaAs/GaAs quantum dots.~\cite{Ohnesorge}

\section{Multiple exciton generation}

As has been demonstrated above, the surface halogenation suppresses both radiative and Auger recombination in nanocrystals. These processes are reverse ones with respect to the carrier multiplication (or multi-exciton generation): they tend to decrease the number of excitons in a system, while the process of multi-exciton generation, shown schematically in Fig. 4, has an opposite trend. Initially, a high-energy photon creates a highly excited electron-hole pair which then reduces its energy creating one more electron-hole pair with lower energy. As a result, two (or, even, more) excitons can arise in the system after absorption of a single photon.

The multi-exciton generation is a basic process for photovoltaics, where light energy transforms into electric current. Its realization in nanocrystals has been experimentally confirmed.~\cite{Klimov,Stolle,Beard,Nozik,Boer,Trinh} In order to be more efficient, this process should be faster than other competitive processes occurring along with exciton generation, such as inter-band radiative recombination or Auger recombination. From this point of view, slowing the latter down caused by surface halogenation is an extremely positive factor. It is now interesting to understand how halogen passivation influences the multi-exciton generation itself.

To this goal, we calculate the rates of the multi-exciton generation in the Si$_{317}$X$_{172}$ crystallite (X = H, Cl, and Br):
\begin{equation}
\tau_{G}^{-1} = \frac{2\Gamma}{\hbar}\sum_{f_1}\sum_{i_2}\sum_{f_2}\frac{|\langle\Psi_{i_1i_2}|{\hat U}|\Psi_{f_1f_2}\rangle|^2}{\Gamma^2+(\varepsilon_{i_1} + \varepsilon_{i_2} - \varepsilon_{f_1} - \varepsilon_{f_2})^2},
\end{equation}
where ${\hat U} = {\hat U}_0 + {\hat U}_1$, as before, while $\Psi_{i_1i_2}$ and $\Psi_{f_1f_2}$ are the products of the single-particle Kohn-Sham wave functions of the initial or final electron states participating in the transition, as shown in Fig. 4. Here, we calculated the rates for the Si$_{317}$X$_{172}$ crystallite within the range of excess energy $0<\Delta\varepsilon<0.6$ eV, where $\Delta\varepsilon = \varepsilon_{i_1} - \varepsilon_{LUMO} - \varepsilon_g$ for the process initiated by highly excited electron, and $\Delta\varepsilon = \varepsilon_{HOMO} - \varepsilon_{f_1} - \varepsilon_g$ for the hole-initiated process. The calculated rates are shown in Fig. 5. It is evident that the rates rise in totality as the excess energy increases, because of the considerable increase in the number of possible states participating in the transitions with increasing $\Delta\varepsilon$, which opens up many new channels for the realization of exciton generation.

\begin{figure}[t]
  \centering
  \includegraphics[scale=0.9]{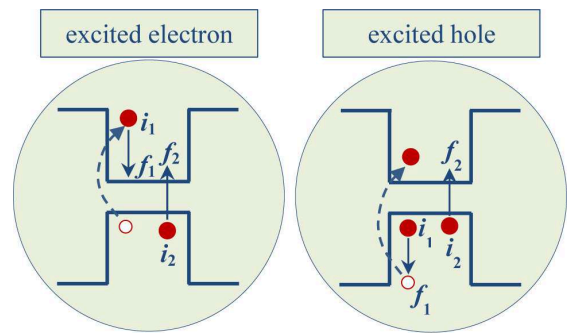}
  \caption{Exciton generation initiated by a highly excited electron (left) or hole (right). Dashed arrows indicate the formation of the initial exciton (electron-hole pair) caused by the absorption of a photon. Coulomb interaction between the electrons in the states $i_1$ and $i_2$ induces an energy exchange resulting in the electron transitions to the final states $f_1$ and $f_2$ indicated by solid arrows. As a result, one more exciton appears in the system.} \label{figure4}
\end{figure}

\begin{figure}[b]
  \centering
  \includegraphics[scale=0.75]{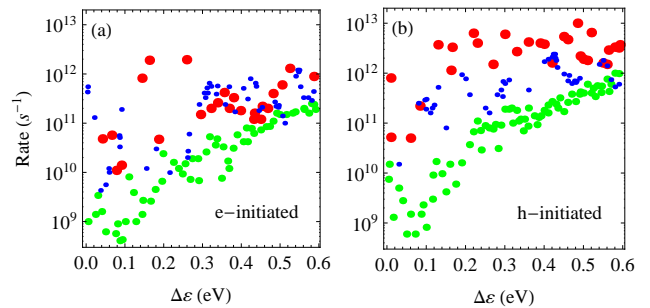}
  \caption{Rates of the exciton generation initiated by highly excited (a) electron, and (b) hole (as shown in Fig. 4) in Si$_{317}$X$_{172}$ crystallite for: X = H (small blue dots); X = Cl (medium green dots); and X = Br (big red dots).} \label{figure5}
\end{figure}

It is important to emphasize that the bromination of Si crystallite increases the exciton generation rates compared to H-passivated crystallite, especially if the process is initiated by a highly excited hole. The rates of exciton generation in chlorinated Si crystallite turn out to be less than those in the hydrogenated one at small excess energies. Meanwhile, upon approaching $\Delta\varepsilon \sim 0.5$ eV, $\tau_{G}^{-1}$ in the Si$_{317}$Cl$_{172}$ crystallite increases and tends to the typical values observed in Si$_{317}$H$_{172}$ crystallite. Accordingly, it is possible to conclude that halogen coating of Si crystallites, at least, does not reduce their ability to generate excitons, particularly in cases where the excess energies are not too small. This is in contrast to radiative and Auger recombination, where the rates became substantially less due to halogenation.

This means that halogenation of Si nanocrystals can increase an efficiency of the photon-to-exciton conversion, which is defined by an excess of the number of created excitons ($n$) over the number of absorbed photons ($N$):~\cite{Klimov,Haverkort} $\eta = n/N > 1$ (internal quantum efficiency). There are many experimental works in which external quantum efficiency was measured in Si crystallites,~\cite{Stolle,Beard,Nozik,Boer,Trinh} as well as in crystallites formed from A$_4$B$_6$, A$_2$B$_6$, or A$_3$B$_5$ semiconductors.~\cite{Klimov,Schaller,Nair,Davis,Hu,Smith} The authors reported on the observation of multi-exciton generation in the investigated systems.

Theoretical consideration of the exciton kinetics in the halogen-coated Si crystallites~\cite{PCCP} has revealed strong dependence of $\eta$ on the quantitative relationship between the rates $\tau_G^{-1}$ and $\tau_A^{-1}$. According to the obtained results the decrease in the Auger rate (caused by the halogenation), and its absence in the multi-exciton generation rate, is accompanied by a gradual increase in the quantum efficiency of the order of tens of percentages.

\section{Energy transfer in ensembles of nanocrystals}

All the considered above processes may occur in isolated nanocrystals. Meanwhile, usually in experiments, as was already pointed out in the Introduction, it is necessary to deal with ensembles of nanocrystals, where non-radiative energy exchange between nanocrystals takes place and strongly influences the ensemble photoluminescence.~\cite{PhysE09,JLumin} Such an energy transfer is occurred owing to the tunnel or F\"{o}rster~\cite{Forster} migration of excitons. Below, we consider both these mechanisms.

\subsection{Exciton tunneling}

In order to estimate an intensity of the tunnel transfer in ensembles of nanocrystals we shall compute the rate of the exciton tunneling from some nanocrystal with radius $R_1$ to an adjacent nanocrystal with greater radius $R_2$. We assume the radii to be close enough, so that $2r = R_2 - R_1 \ll 2R = R_1 + R_2$, which means an approximate equality of the exciton energies in both crystallites. Initial state of the system corresponds to the presence of exciton with the lowest possible energy $\varepsilon_g(R_1)$ in the first nanocrystal and its absence in the second one, and \textit{vice versa} for the final state. In the considered case ($r\ll R$) the rate of such a transition can be presented in the following form:~\cite{Reich}
\begin{equation}
\tau_{T}^{-1} = \left(\frac{T_eT_h}{\varepsilon_{ee}}\right)^2\frac{8\Gamma/\hbar}{\Gamma^2+(\varepsilon_g(R_1)-\varepsilon_g(R_2))^2}
\end{equation}
where $T_e$ and $T_h$ are the single-particle electron and hole tunnel matrix elements,~\cite{JAP} respectively, and $\varepsilon_{ee}\approx e^2/2\epsilon_{out}R$ is the energy of the intermediate states equal to the Coulomb energy of the charged nanocrystal. The two possible intermediate states correspond to the first or the second charged nanocrystals with the charges $\pm e$ or $\mp e$. In order to estimate the tunnel rate numerically we set $\Gamma = 10$ meV, inter-crystallite distance $L =0$, and consider exactly resonant tunneling, i.e. $R_1 = R_2 = R$ and, correspondingly, $\varepsilon_g(R_1) = \varepsilon_g(R_2)$. This yields, in fact, highest possible values. If $L$ or $r$ (or both) becomes nonzero and rises, the rate sharply decreases and tends to zero.

The exciton tunnel rate for this case is shown in Fig. 6 as function of $R$. The obtained dependence was calculated for silicon nanocrystals, but it will be qualitatively similar for nanocrystals of direct-band-gap semiconductors, because tunnel transitions occur within same bands, and the gap type (direct or indirect) does not, in fact, affect the process.

\begin{figure}
  \centering
  \includegraphics[scale=0.9]{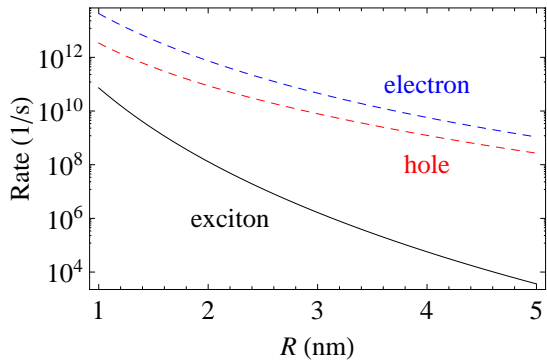}
  \caption{Tunneling rates as functions of the nanocrystal radius for excitons, electrons, and holes in equivalent Si nanocrystals contacting each other.} \label{figure6}
\end{figure}

As shown in the figure, the exciton tunnel rate equals $\sim 10^{10}$ s$^{-1}$ for $R=1$ nm and gradually drops to $\sim 10^4$ s$^{-1}$ for $R=5$ nm. Thus, for crystallites formed of direct-band-gap semiconductors, highest possible rates of exciton tunneling (at $R=1$ nm) turn out to be of the order of typical rates of radiative recombination, although the rates of the single-particle (electron or hole) tunnel migration can exceed the latter. For instance, for 1 nm-radius nanocrystals, the tunnel rates of the electrons and holes exceed $10^{13}$ s$^{-1}$ and $10^{12}$ s$^{-1}$, respectively.~\cite{JAP} In silicon crystallites, the exciton tunnel transitions are substantially faster than the radiative ones within the whole range of radii considered here.

We should emphasize that such high values of the tunnel rate take place only in the resonant case and if the crystallites touch each other ($L = 0$). As inter-crystallite distance $L$ increases, the rate of exciton tunneling drops as $\exp[-2(q_e+q_h)L]$, where $q_e\sim q_h$ is the electron, or the hole, wave-function decrement in a barrier, while individual tunnel rates of electrons and holes behave as $\exp(-2q_{e,h}L)$ only. Thus, exciton tunneling has much less radius of action than the electron or hole tunneling.

It should be noted, however, that the single-particle tunneling can be strongly suppressed by the Coulomb blockade effect that has to be taken into account when calculating the individual tunnel rates. At the same time, the exciton tunneling does not accompanied by the charge transfer and, correspondingly, the nanocrystal charging. As a result, the Coulomb blockade does not, in fact, affect the tunnel migration of excitons.

\subsection{Resonant exciton transfer}

The resonant exciton transfer is mainly realized through the electrostatic F\"{o}rster interaction of two dipoles in two adjacent crystallites:~\cite{Allan}
\begin{equation}
V({\bf r}_1,{\bf r}_2) = \frac{\kappa e^2}{\epsilon_{out}b^3}\left[{\bf r}_1{\bf r}_2 - 3\frac{({\bf r}_1{\bf b})({\bf r}_2{\bf b})}{b^2} \right],
\end{equation}
where ${\bf b}$ is the inter-crystallite center-to-center vector. In one crystallite, the electron-hole pair annihilates and transfers its energy into a neighboring crystallite where a new electron-hole pair is excited. Thus, virtual transfer of excitons between two crystallites can be realized without charge transfer.

In order to calculate the rate of the F\"{o}rster exciton transition from a crystallite with a radius $R_1$ into a neighboring crystallite with a radius $R_2$, we use, as before, the Fermi golden rule:
\begin{equation}
\tau_{F}^{-1} = \frac{2\Gamma}{\hbar}\frac{|\langle\Psi_i|{\hat V}|\Psi_f\rangle|^2}{\Gamma^2 + (\varepsilon_g(R_1) - \varepsilon_g(R_2))^2}.
\end{equation}
Here, ${\hat V}$ is the dipole interaction operator, $\Psi_i = \psi_c({\bf r}_1)\psi_v({\bf r}_2)$ is the wave function of the initial two-particle state with the energy $\varepsilon_g(R_1)$ coinciding with the energy gap of the first crystallite.
Initially, in the conduction band of the first crystallite, one electron exists with the wave function $\psi_c({\bf r}_1)$ describing the ground state, while in the valence band a hole is created. In the second crystallite, electron occupies the ground valence state described by the wave function $\psi_v({\bf r}_2)$. Due to the dipole interaction the electron-hole pair (exciton) in the first crystallite nonradiatively annihilates transferring its energy into the second crystallite where new exciton with the energy $\varepsilon_g(R_2)$ is created. In the final state, the system has the wave function $\Psi_f = \psi_v({\bf r}_1)\psi_c({\bf r}_2)$.


In ensembles formed of direct-band-gap A$_2$B$_6$ or A$_3$B$_5$ semiconductor crystallites, the exciton transfer has the rates $\sim 10^9 - 10^{10}$ s$^{-1}$,~\cite{Kagan,Crooker,Allan,Scholes,Lunz,Lyo} which are of the same order of magnitude as the radiative recombination rates in these nanocrystals. In ensembles of silicon crystallites the exciton transfer turns out to be much slower than the radiative recombination: its rates are two to three orders of magnitude less~\cite{Allan,JCTN,Gusev} (not greater than $\sim 10^3$ s$^{-1}$). Doping of silicon nanocrystals with phosphorus allows to increase the rates up to the values comparable with those of radiative recombination~\cite{PRB2} ($\lesssim 10^8$ s$^{-1}$ for nanocrystal radius $R\gtrsim 1$ nm). Nevertheless, these values remain still smaller than the rate values for the direct-gap semiconductor nanocrystals.

\section{Concluding remarks}

Summarizing all the aforesaid, we would like to emphasize, once again, that the radiative properties and efficiency of light emission in semiconductor nanostructures is determined not only by the radiative transitions but also by various non-radiative processes which can be even more intensive than the radiative transitions themselves. Usually, and especially, this turns out to be important for the nanostructures of indirect-band-gap materials, such as, e.g., ensembles of silicon nanocrystals, where migration processes and multi-exciton transitions play a dominant role and suppress the radiative recombination in a significant extent. It was found,~\cite{JLumin} in particular, that the inter-crystallite migration of excitons in the ensemble causes quenching the luminescence from small Si nanocrystals, which leads to the red-shifting and narrowing the luminescence peak. On the other hand, even in silicon nanocrystals, the non-radiative processes can be used in a constructive way. For instance, the exciton migration is accompanied by a non-radiative energy transfer that can be made a directed one \textit{via} creating a certain ``architecture'' of the nanocrystals in the ensemble, which allows to concentrate and illuminate the energy inside a given area.~\cite{Kawazoe,JCTN} The multi-exciton effects make it possible to efficiently transform the absorbed photons into the rising number of electron-hole pairs capable of participating in the electric current. Such a transformation underlies an operating principle of solar cells.

\section*{Acknowledgments}

The work was supported by the Ministry of Science and Higher Education of the Russian Federation (Assignment No 3.2637.2017/4.6).

\end{document}